\documentclass[reqno]{amsart}

\usepackage[usenames,dvipsnames,svgnames,table]{xcolor}
\usepackage{amssymb,latexsym,amsfonts,amsmath}
\usepackage{graphicx}
\usepackage{eso-pic}
\usepackage{epsfig}
\usepackage{cuted}
\usepackage{mathtools}
\usepackage{tikz}
\usetikzlibrary{automata}
\usetikzlibrary{shapes}
\usetikzlibrary{calc,shapes,arrows} 
\usetikzlibrary{arrows.meta}
\usepackage{mdwlist}
\usepackage{refcount}
\usepackage{comment}
\usepackage{multicol}
\usepackage{carshapes}
\usepackage{geometry}

\usepackage{cite}

\usepackage{enumitem}

\usepackage{keyval,trig}
\usepackage{amssymb,dsfont}
\usepackage{mathtools,mathrsfs}

\usepackage{hyperref}
\hypersetup{colorlinks=true,linkcolor=blue,citecolor=blue,filecolor=blue,urlcolor=blue}

\usepackage[linesnumbered,ruled,resetcount]{algorithm2e}
\usepackage{etoolbox}
\newcommand{\algrule}[1][.2pt]{\par\vskip.5\baselineskip\hrule height #1\par\vskip.5\baselineskip}

\newcounter{algoline}

\AtBeginEnvironment{algorithm}{\setcounter{algoline}{0}}
\usepackage{ragged2e}

\usepackage{tikz}
\usetikzlibrary{calc,shapes,arrows}

\usepackage{xspace}
\usepackage{caption}
\usepackage{subcaption}
\usepackage{multirow, booktabs, array, tabularx}

\newtheorem{theorem}{Theorem}[section]

\newtheorem{problem}[theorem]{Problem}
\newtheorem{proposition}[theorem]{Proposition}

\newtheorem{definition}[theorem]{Definition}

\newtheorem{remark}[theorem]{Remark}

\usepackage[many]{tcolorbox}
\usetikzlibrary{calc}
\tcbuselibrary{skins}
\definecolor{forestgreen(web)}{rgb}{0.13, 0.55, 0.13}
\definecolor{deepskyblue}{rgb}{0.0, 0.75, 1.0}
\definecolor{flame}{rgb}{0.89, 0.35, 0.13}
\definecolor{brilliantrose}{rgb}{1.0, 0.33, 0.64}
\definecolor{chestnut}{rgb}{0, 0.4470, 0.7410}
\definecolor{darklavender}{rgb}{0.45, 0.31, 0.59}
\definecolor{darktangerine}{rgb}{1.0, 0.66, 0.07}
\definecolor{khakestari}{RGB}{205,205,205}
\newtcolorbox{resp}[1][]{%
enhanced jigsaw,%
colback=gray!5!white,%
colframe=gray!80!black,%
size=small,%
boxrule=1pt,%
halign title=flush center,%
coltitle=black,%
breakable,%
drop shadow=black!50!white,%
attach boxed title to top left={xshift=1cm,yshift=-\tcboxedtitleheight/2,yshifttext=-\tcboxedtitleheight/2},%
minipage boxed title=3cm,%
boxed title style={%
	colback=white,%
	size=fbox,%
	boxrule=1pt,%
	boxsep=2pt,%
	underlay={%
		\coordinate (dotA) at ($(interior.west) + (-0.5pt,0)$);
		\coordinate (dotB) at ($(interior.east) + (0.5pt,0)$);
		\begin{scope}[gray!80!black]
			\fill (dotA) circle (2pt);
			\fill (dotB) circle (2pt);
		\end{scope}
	}%
},%
#1%
}

\xdefinecolor{MATLABblue}{rgb}{0.0,0.45,.74}
\xdefinecolor{MATLABred}{rgb}{0.85,0.33,.1}

\newcommand{\sv}{{\texttt{sub}}}
\newcommand{\cv}{{\texttt{col}}}
\newcommand{\R}{{\mathbb{R}}}

\newcommand{\N}{{\mathbb{N}}}

\newcommand{\ie}{{\it i.e.}}
\newcommand{\eg}{{\it e.g.}}

\usepackage[normalem]{ulem}

\newcommand{\Let}{:=}

\definecolor{fluorescentpink}{rgb}{1.0, 0.08, 0.58}

\usepackage{fancyhdr}

\newenvironment{nouppercase}{%
	\renewcommand{\uppercasenonmath}[1]{}}{}

\begin{document}

\begin{abstract}
This paper offers a formal framework for the rare collision risk estimation of autonomous vehicles (AVs) with multi-agent situation awareness, affected by different sources of noise in a complex dynamic environment. In our proposed setting, the situation awareness is considered for one of the ego vehicles by aggregating a range of diverse information gathered from other vehicles into a vector. We model AVs equipped with the situation awareness as \emph{general stochastic hybrid systems} (GSHS) and assess the probability of collision in a \emph{lane-change scenario} where two self-driving vehicles simultaneously intend to switch lanes into a shared one, while utilizing the \emph{time-to-collision} measure for decision-making as required. Due to the substantial data requirements of simulation-based methods for the rare collision risk estimation, we leverage a \emph{multi-level} importance splitting technique, known as \emph{interacting particle system-based estimation with fixed assignment splitting} (IPS-FAS). This approach allows us to estimate the probability of a rare event by employing a group of interacting particles. Specifically, each particle embodies a system trajectory and engages with others through resampling and branching, focusing computational resources on trajectories with the highest probability of encountering the rare event. The effectiveness of our proposed approach is demonstrated through an extensive simulation of a lane-change scenario.\\

\end{abstract}

\title{\LARGE Rare Collision Risk Estimation of Autonomous Vehicles with Multi-Agent Situation Awareness}

\author{{\bf {\large Mahdieh Zaker$^1$, Henk A.P. Blom$^2$, Sadegh Soudjani$^3$, and Abolfazl Lavaei$^1$}}\\{\normalfont $^1$School of Computing, Newcastle University, United Kingdom}\\\textsf{\{m.zaker2,abolfazl.lavaei\}@newcastle.ac.uk}\\{\normalfont $^2$Delft University of Technology, The Netherlands}\\ \textsf{h.a.p.blom@tudelft.nl}\\{\normalfont $^3$Max Planck Institute for Software Systems, Germany}\\\textsf{sadegh@mpi-sws.org}}

\pagestyle{fancy}
\lhead{}
\rhead{}
  \fancyhead[OL]{M. Zaker, H. A.P. Blom, S. Soudjani, and A. Lavaei}

  \fancyhead[EL]{\textbf{Rare Collision Risk Estimation of Autonomous Vehicles with Situation Awareness}}
  \rhead{\thepage}
 \cfoot{}
 
\begin{nouppercase}
	\maketitle
\end{nouppercase}

\section{Introduction}\label{sec:intro}
Autonomous vehicles (AVs) have been becoming increasingly popular on a daily basis, primarily due to their numerous advantages, including the reduction of air pollution, alleviation of traffic congestion, and mitigation of human-error-related fatalities. However, these complex systems operate within dynamic environments where they interact with a diverse range of factors, presenting various uncertainties, such as unpredictable weather conditions, unexpected pedestrian movements, and the wide-ranging driving behaviors of human operators. Given that accidents in this context can pose significant risks to human safety, AVs are considered safety-critical systems~\cite{knight2002safety} and ensuring their safe operation in complex and uncertain environments is a paramount challenge that demands significant attention.

Improving the safety of AVs and reducing their collision risks involve leveraging information from all agents to enhance AVs awareness. By doing so, AVs can, in specific scenarios, make informed decisions based on the concept of \emph{situation awareness} (SA) \cite{wardzinski2008safety}, by involving the knowledge of ongoing events \cite{wardzinski2006role}. This strategic approach is capable of significantly decreasing collision risks, even to the extent of making them \emph{exceedingly rare}, such as less than $10^{-7}$. Consequently, it becomes imperative to  investigate the impact of situation awareness in such challenging scenarios, when the probability of AV rare collisions  approaches zero ($\approx 0$). When dealing with rare events, typically characterized by a probability less than $10^{-7}$, \textit{Monte Carlo} methods \cite{metropolis1949monte} become unfeasible, unless an impractically large sample size is utilized. In such situations, alternative methods are sought to accurately estimate the probability of rare events within a reasonable sample frame.

There have been various approaches proposed for estimating collision risk of AVs. Existing results include \textit{importance sampling} \cite{bucklew2004introduction}, which involves selecting a sampling distribution and weighting samples by the likelihood ratio between sampling and target distributions. However, finding an appropriate sampling distribution can be challenging, and as the problem dimensions increase, the likelihood ratio becomes less reliable, leading to its avoidance in high-dimensional problems \cite{botev2008efficient}. Another alternative method, known as \emph{importance splitting} \cite{glasserman1999multilevel} (also referred to as multi-level splitting, splitting, or subset simulation), aims to overcome these challenges. In particular, importance splitting approach treats rare events as nested occurrences with relatively higher probabilities, focusing on propagating realizations that are likely to lead to the rare event (mutation phase), while discarding others (selection phase) (see \eg, \cite{l2009splitting, morio2015estimation}).

The \textit{interacting particle system} (IPS) \cite{del2005genealogical, cerou2006genetic} is analogous to an importance sampling algorithm, with the distinction that the interacting particles possess resampling weights. This relatively recent algorithm is founded on evolutionary principles and enhances the estimation of rare-event probabilities. Unlike the importance sampling method, where both selection and mutation stages are applied to the entire Markov trajectory, IPS applies these stages at various times during the evolution of the Markov process. This distinction makes IPS a potentially more attractive option compared to the importance sampling method~\cite{morio2013optimisation}, particularly when simulating entire Markov trajectories is time-consuming (see \eg,  \cite{blom2006particle, blom2007free,MA2023101303,bujorianu2006toward}). 

There have also been findings regarding situation awareness as a method to enhance system safety. Existing results include situation awareness for dynamic systems, initially introduced in \cite{endsley1995toward}, which encompasses perceiving environmental elements across time and space; a formal framework based on the concept of multi-agent SA (MA-SA) relations in a system of multiple agents \cite{blom2015modelling}; AV safety assurance, with a particular emphasis on dynamic risk assessment (DRA)~\cite{wardzinski2008safety}; and a model-based framework based on SINADRA for dynamic risk assessment that balances residual risk and driving performance \cite{reich2020sinadra}. In complex multi-agent scenarios, a \emph{compositional data-driven} approach for formally estimating collision risks of AVs with black-box dynamics is introduced in \cite{lavaei2022formal}.

\begin{figure}[t!]
	\centering 
	\includegraphics[width=0.87\linewidth]{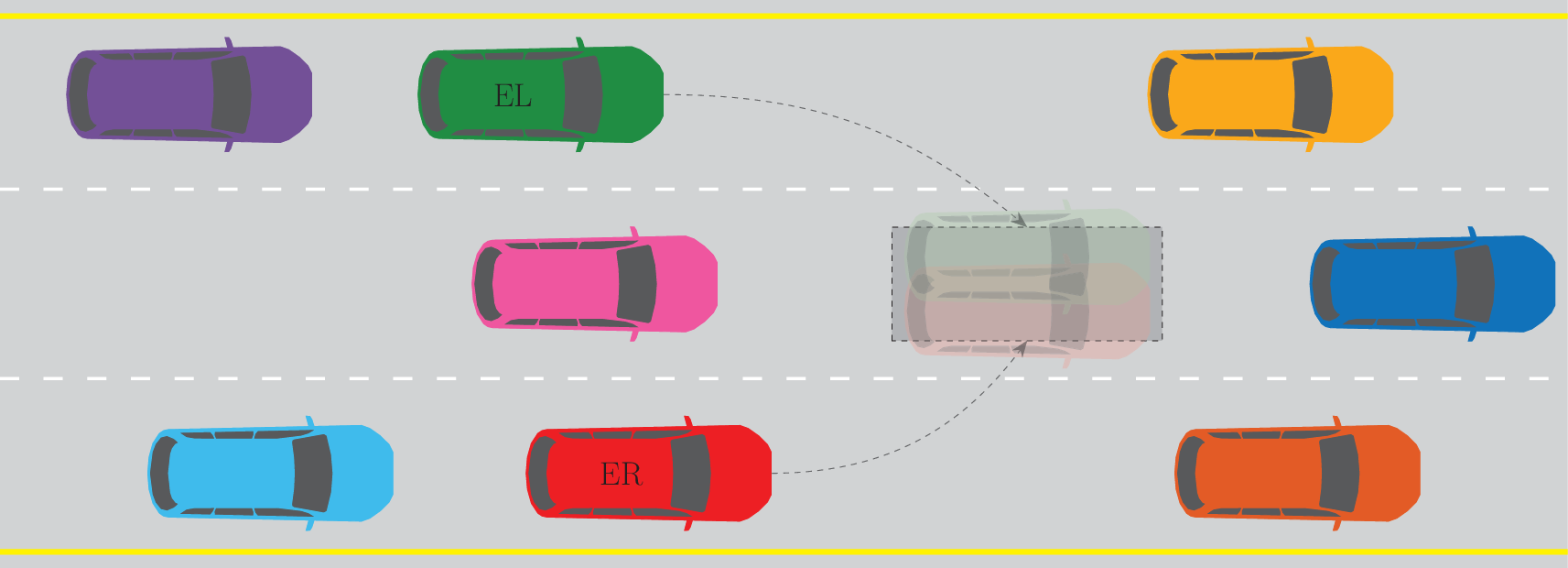}
	\caption{Lane-change scenario: Two AVs are recognized as $e \in \mathcal{E} = \{ EL, ER \}$, with the first letter indicating their status as ego vehicles and the second letter specifying their lane (right or left).} 
	\label{fig:3lane}
\end{figure}

{\bf Original contributions}. This paper introduces a formal approach for the rare collision risk estimation of AVs operating on a three-lane road alongside human-driven vehicles, by utilizing the interacting particle system-based estimation with fixed assignment splitting (IPS-FAS) algorithm~\cite{MA2023101303}. We model each AV as a general stochastic hybrid system (GSHS) to capture various sources of noise and uncertainty. The primary scenario under examination is a lane-change situation, in which AVs are positioned in the first and third lanes, each followed by a human-driven vehicle, with an unoccupied space in the second lane. At a specific time instant, both AVs make a decision to change lanes.
When the AV with situation awareness detects the other AV's lane-change intention, it computes the time-to-collision \cite{greene2011efficient} and utilizes this measurement to determine whether to proceed with its lane change or revert to its original lane. Our primary objective is to compute the potentially \emph{rare collision probability} for these two AVs under different conditions, incorporating situation awareness and computing the time-to-collision measure. The scenario under investigation is visually represented in Fig. \ref{fig:3lane}, where the ego AVs are depicted as the red and green vehicles.

\section{General Stochastic Hybrid Systems}

\subsection{Preliminaries and Notation}
We denote the set of real and positive real numbers by $\R$ and $\R^+$, respectively, while $\N \Let \{ 1, 2,\dots\}$ represents the set of positive integers. The logical AND and OR operations are denoted by $\wedge$ and $\vee$, respectively. Symbols $\pmb{1}_{n\times m}$ and $\pmb{0}_{n\times m}$ are matrices of $n\times m$ dimension, consisting of unit and zero elements, respectively. We denote the empty set by $\emptyset$. We use $\operatorname{col}(\cdot)$ to create a vector from its input arguments. Moreover, we denote the exponential distribution with rate parameter $\lambda$ by $\exp(\lambda)$. For a sequence of scalars $(\gamma_1, \gamma_2, \dots, \gamma_n)$, we use the notation $\prod_{k=1}^{n} \gamma_k$ to represent the product $\gamma_1 \cdot \gamma_2 \cdot \dots \cdot \gamma_n$. 
The system's state in this work is hybrid, represented by both a continuous variable, denoted as $\mathbf{x}$, and a discrete variable, denoted as $\theta$. The continuous variable evolves in some open sets in Euclidean space $X^\theta$, while the discrete variable is an element of a countable set $\Theta$. The hybrid state space is denoted by $\Xi \triangleq \bigcup_{\theta \in \Theta} \{ \theta \} \times X^\theta$, and $\overline{\Xi} = \Xi \cup \partial\Xi$ represents the closure of $\Xi$, where $\partial\Xi \triangleq \bigcup_{\theta \in \Theta} \{ \theta \} \times \partial X^\theta$ is the boundary of $\Xi$.

We consider a probability space $(\Omega, \mathcal{F}_{\Omega}, \mathds {P}_{\Omega})$, where $\Omega$ is the sample space, $\mathcal{F}_{\Omega}$ is a $\sigma$-algebra on $\Omega$ comprising subsets of $\Omega$ as events, and $\mathds{P}_{\Omega}$ is a probability measure that assigns probabilities to events. We assume that triple $(\Omega, \mathcal{F}_{\Omega}, \mathds{P}_{\Omega})$ is endowed with a filtration $\mathbb{F}=(\mathcal{F}_s)_{s \geq 0}$ satisfying the usual conditions of completeness and right continuity. Let $\left(\mathbb{W}_s\right)_{s \geq 0}$ be an
$m$-dimensional $\mathbb{F}$-Brownian motion, and $\left(\mathbb{P}_s\right)_{s \geq 0}$ be an $m'$-dimensional $\mathbb{F}$-Poisson process (mutually independent). If $X$ is a Hausdorff topological space, $\mathcal{B}(X)$ denotes its Borel $\sigma$-algebra, and $(X, \mathcal{B}(X))$ is a Borel space. 

\subsection{General Stochastic Hybrid Systems}

As autonomous vehicles operate in a complex environment interacting with various entities, they might face numerous unpredicted events. Henceforth, different sources of uncertainty should be taken into account when modeling AVs. To do so, we employ the notion of general stochastic hybrid systems, which encompasses a wide range of stochastic phenomena,
as follows~\cite{bujorianu2006toward}.
\begin{definition}[GSHS]\label{def:GSHS}
	Each agent of AVs is modeled as a general stochastic hybrid system (GSHS), denoted by $\mathcal{A} = \big(( \Theta , d, \mathcal{X}), f, g, Init, \lambda, R\big)$, where
	\begin{itemize}
		\item $\Theta$ is a countable set of discrete variables;
		
		\item $d:\Theta \rightarrow \N$ is a mapping that provides the dimensions of the continuous state spaces for each element in $\Theta$;
		
		\item $\mathcal{X}:\Theta \rightarrow \R^{d(\cdot)}$ associates each $\theta \in \Theta$ with an open subset $X^\theta$ within $\R^{d(\theta)}$;
		
		\item $f : \Xi \rightarrow \R^{d(\cdot)}$ is a vector field;
		
		\item $g: \Xi \rightarrow \R^{d(\cdot) \times m}$ is an $X^{(\cdot)}$-valued matrix, $m \in \N$;
		
		\item $Init : \mathcal{B}(\Xi) \rightarrow [0, 1]$ is an initial probability measure on $(\Xi, \mathcal{B}(\Xi))$;
		
		\item $\lambda: \overline{\Xi}\rightarrow\R^+$ is a transition rate function;
		
		\item $R:\overline{\Xi}\times\mathcal{B}(\overline{\Xi})\rightarrow [0,1]$ is a transition measure.
	\end{itemize}
	Then, a stochastic process $\{ \theta_t, \mathbf{x}_t \}$ is called a GSHS execution if there exists a sequence of stopping times $s_0 = 0 < s_1 < s_2 < \cdots$ such that for each $j\in \N$,
	\begin{itemize}
		\item $(\theta_0, \mathbf{x}_0)$ is a $\Xi$-valued random variable extracted according to the probability measure $Init$;
		
		\item For $t\in [s_{j-1}, s_j)$, $\theta_t$, $\mathbf{x}_t$ is a solution of the stochastic differential equation (SDE):
		\begin{align*}
		\begin{cases}
			\mathsf{d}\theta_t = 0\\
			\mathsf{d}\mathbf{x}_t=f(\theta_t, \mathbf{x}_t)\mathsf{d}t + g(\theta_t, \mathbf{x}_t)\mathsf{d}\mathbb{W}_t
		\end{cases}
		\end{align*}
		in which $\mathbb{W}_t$ is an $m$-dimensional standard Brownian motion. 
		\item $s_j$ is the minimum of the following two stopping times:
		\begin{enumerate}[label=(\roman*)]
			\item first hitting time $t > s_{j-1}$ of the boundary of $X^{\theta_{s_{j-1}}}$ by the phase process $\{ \mathbf{x}_t \}$;
			\item first moment $t > s_{j-1}$ of a transition event to happen at rate $\lambda(\theta_t, \mathbf{x}_t)$.
		\end{enumerate}
		
		\item At the stopping time $s_j$ the 
		hybrid state $( \theta_{s_j}, \mathbf{x}_{s_j} )$ meets the conditional probability measure $p_{\theta_{s_j},\mathbf{x}_{s_j}|\theta_{s_j-}, \mathbf{x}_{s_j-}}(A|\theta,\mathbf{x})=R((\theta, \mathbf{x}), A)$ for all $A\in \mathcal{B}(\Xi)$, where $s_j-$ indicates the time instant immediately before the stopping time $s_j$ is reached. 
	\end{itemize}
\end{definition}

According to \cite{MA2023101303}, a GSHS can be transformed to an SHS, as a \emph{more tractable model}, involving four key modifications:
\begin{enumerate}[label=(\roman*)]
	\item An auxiliary state component $q_t$, representing ``remaining local time'', is initialized at a specific stopping time $\tau$ with an initial condition of $q_\tau \sim \exp(1)$;
	\item The exit boundary of $X^\theta$ is expanded by introducing an additional boundary condition, where $q_{t-} = 0$, \emph{i.e.,} the value of $q_t$ just before $t$;
	\item Spontaneous probabilistic jumps in $\{\theta_t, \mathbf{x}_t\}$ are replaced by forced probabilistic jumps occurring at the moment when $q_{t-} = 0$;
	\item When the extended exit boundary is reached at the stopping time $\tau'$, the ``remaining local time'' is resampled as $q_{\tau'}\sim \exp(1)$.
\end{enumerate}
Subsequently, the GSHS $\mathcal{A} = \big(( \Theta , d, \mathcal{X}), f, g, Init, \lambda, R\big)$ is transformed to the SHS $\mathcal{A}^\ast = \big(( \Theta^\ast ,$ $ d^\ast, \mathcal{X}^\ast), f^\ast, g^\ast, Init^\ast, R^\ast\big)$ as follows:
\begin{itemize}
	\item $\Theta^\ast = \Theta,$ $d^\ast = d+1$, $\mathcal{X}^\ast = \mathcal{X}\times (0,\infty)$;
	
	\item $f^\ast(\theta_t,\mathbf{x}_t,\cdot) = \big[f(\theta_t, \mathbf{x}_t)\quad -\lambda(\theta_t, \mathbf{x}_t)\big]^\top$;
	
	\item $g^\ast(\theta_t, \mathbf{x}_t, \cdot) = \big[g(\theta_t,\mathbf{x}_t)\quad 0\big]^\top$;
	
	\item $Init^\ast = \big[Init\quad q_0\big]^\top$ with $q_0\sim \exp(1)$;
	
	\item $R^\ast\big((\theta_t, \mathbf{x}_t, \cdot);A\times \mathsf{d}q\big) = R\big((\theta_t, \mathbf{x}_t);A\big)\times e^{-q}\mathsf{d}q$.
\end{itemize}

This transformation is mainly helpful for system execution (cf. Algorithm~\ref{alg:Algorithm 1}), upon which one can estimate the rare event probability (cf. Algorithm~\ref{alg:Algorithm 2}). Having delved into the general stochastic hybrid system for modeling AVs, in pursuit of our goal to enhance the safety of AVs in the lane-change scenario, we leverage a \emph{multi-agent situation awareness} framework in the following section.

\section{Multi-Agent Situation Awareness}

Here, we leverage the concept of multi-agent situation awareness (MA-SA), building upon the fundamental work in \cite{blom2015modelling}. In a multi-agent system of $N$ agents $\mathcal{A}_i,\; i \in \{1,\dots, N\}$, each agent 
has state $z_{t,i}$ at time instant $t$, comprising of SA and non-SA states.
The multi-agent situation awareness relation of agent $\mathcal{A}_i$ regarding agent $\mathcal{A}_j$ is represented by $Z_i^j$, which is a set of $N_i^j$ different pairs $(s,r)_n,\;n\in\{1,\dots, N_i^j\}$ such that $s$ references state element $z_{t,i}(s)$ and $r$ references state element $z_{t,j}(r)$. Subsequently, the SA of agent $\mathcal{A}_i$ about the state of agent $\mathcal{A}_j$ at time instant $t$ is represented by $\sigma_{t,i}^j$ as follows:
\begin{equation}\label{eq:SA_frame}
	\sigma_{t,i}^j\triangleq\big\{ z_{t,i}(s),\; \exists r\; \text{s.t.}\; (s, r) \in Z^j_i\big\}\!.
\end{equation}
It can be concluded that non-empty $Z_i^j$ leads to non-empty $\sigma_{t,i}^j$, which means agent $\mathcal{A}_i$ possesses SA about agent $\mathcal{A}_j$.
In addition to the MA-SA components $\sigma_{t,i}^j,\;j\neq i$, $\zeta_{t,i}$ denotes \emph{base state} of $\mathcal{A}_i$, determining state elements of $z_{t,i}$ that are not in relation with any other state element through $\{ Z^j_i,\; j = 1,\dots, N\}$, \ie,
\begin{equation}\label{eq:NonSA_frame}
	\zeta_{t,i}\triangleq\big\{ z_{t,i}(s),\; \text{s.t.}\; (s, r)\notin Z^j_i\; \text{for}\; \forall (j, r) \big\}.
\end{equation}
Following \eqref{eq:SA_frame}-\eqref{eq:NonSA_frame}, the state $z_{t, i}$ of agent $\mathcal{A}_i$ contains base state $\zeta_{t,i}$ and SA of other agents $\sigma_{t,i}^j,\; j\neq i$, as
\begin{equation}
	z_{t,i} = \zeta_{t,i}\bigcup_{j\neq i}\sigma^j_{t,i}.
\end{equation}

Now the problem we aim to address can be formally defined as follows.

\begin{tcolorbox}
	\begin{problem}\label{prob: Rare-event-est}
		Consider two AVs $i,j\in\mathcal{E},$ driving in the first and third lanes of a three-lane highway, each is followed by another vehicle while also following a leading one. There is a free spot in the second lane, as illustrated in Fig. \ref{fig:3lane}, and there exists a specific moment in time when both vehicles decide to change lanes. Quantify the \emph{potentially rare collision probability} $\gamma$ of these two subject vehicles by assuming that $ER$ possesses SA and is modeled as a GSHS $\mathcal{A} = (( \Theta , d, \mathcal{X}), f, g, Init, \lambda, R)$.
	\end{problem}
\end{tcolorbox}

To address Problem~\ref{prob: Rare-event-est}, we detail our underlying framework in the following sections.

\section{IPS-based Rare-event Estimation}\label{sec:IPS}
Here, we estimate the probability $\gamma$ of the hybrid system states $(\theta_t, \mathbf{x}_t )$, reaching a closed subset $D \subset \Xi$ within a finite time interval $[0, T]$, defined as
\begin{equation}
	\gamma = \mathds{P}(\tau < T),
\end{equation}
where $\tau$ is the first time  that $\{ \theta_t, \mathbf{x}_t\}$ enters the set $D$, \ie,
\begin{equation}
	\tau = \inf\{ t > 0, (\theta_t, \mathbf{x}_t)\in D\}.
\end{equation}
The approach to factorizing the reach probability, denoted as $\gamma$, involves the introduction of a sequence denoted as $D_k, k \in\{ 0, \ldots, m\}$, comprising nested closed subsets within the domain $\Xi$. More precisely, we define $D = D_m \subset D_{m-1} \subset \cdots \subset D_1 \subset D_0 = \Xi$, with the specific condition that $D_1$ is chosen to ensure $\mathds{P}\{(\theta_0, \mathbf{x}_0)\in D_1\} = 0$. Furthermore, in order to represent the first time instant at which the pair $(\theta_t, \mathbf{x}_t)$ enters the region $D_k$, $\tau_k$ is defined as 
\begin{equation}
	\tau_k = \inf\{ t > 0 ; (\theta_t, \mathbf{x}_t)\in D_k \vee t \geq T \}.
\end{equation}
To attain the desired factorization, we employ $\{0,1\}$-valued random variables $\chi_k, \;k \in\{0,\dots,m\}$, defined as
\begin{equation}
	\chi_k = \begin{cases}
		1, &~~~ \text{if } \tau_k < T,\\
		0, &~~~ \text{otherwise.}
	\end{cases}
\end{equation}
The factorization presented in the following proposition holds significant practical value by which the reach probability $\gamma$ is expressed as a product of individual probabilities $\gamma_k$. This factorization allows us to systematically explore and estimate the contribution of each level $D_k$ to the overall rare-event probability.
\begin{proposition}
	The factorization is satisfied by the reach probability
	\begin{align}\label{eq:gamma}
		\gamma = \prod_{k=1}^{m}\gamma_k,
	\end{align}
	where $\gamma_k \triangleq \mathbb{E} \big\{\chi_k \!=\! 1 \,\big| \,\chi_{k-1}\!=\!1\big\}=\mathds{P}\big(\tau_k \!<\! T \,\big| \, \tau_{k-1}\!<\!T\big)$.
\end{proposition}
By using the strong Markov property of $\{\theta_t, \mathbf{x}_t\}$, one can develop a recursive estimation of $\gamma$ using the factorization in \eqref{eq:gamma} with $\Xi'\triangleq\R\times\Xi$, $\xi_{k}\triangleq(\tau_k,\theta_{\tau_k}, \mathbf{x}_{\tau_k})$, $Q_k\triangleq(0,T)\times D_k$, for $k \in\{1, \dots,m\}$, and the conditional probability measure $\pi_k(B)\triangleq \mathds{P}(\xi_k\in B|\xi_k\in Q_k)$, for an arbitrary Borel set $B$ of $\Xi'$. A solution to the recursion of transformations is given by $\pi_k$ as follows~\cite{cerou2006genetic}:
\begin{align*}
	\begin{array}{ll}
		\pi_{k-1}(\cdot) \xrightarrow{\text{I. mutation}} & \hspace{-0.25cm}p_k(\cdot)\xrightarrow{\text{III. selection}}\pi_k(\cdot)\\
		& \hspace{-0.25cm}\Big\downarrow\, \text{\footnotesize II. conditioning}\\
		&  \hspace{-0.25cm}\gamma_k
	\end{array}
\end{align*}
where $p_k(B)\triangleq\mathds{P}(\xi_k\in B|\xi_{k-1}\in Q_{k-1})$. By employing the same approach, the following algorithmic steps outline the numerical estimation of $\gamma$ using the IPS method: 
\begin{align*}
	\begin{array}{ll}
		\overline{\pi}_{k-1}(\cdot) \xrightarrow{\text{I. mutation}} & \hspace{-0.25cm}\overline{p}_k(\cdot)\xrightarrow{\text{III. selection}}\Tilde{\pi}_k(\cdot)\xrightarrow{\text{IV. splitting}}\overline{\pi}_k(\cdot)\\
		& \hspace{-0.25cm}\Big\downarrow\, \text{\footnotesize II. conditioning}\\
		&  \hspace{-0.25cm}\overline{\gamma}_k
	\end{array}
\end{align*}
Here, $\overline{\gamma}_k$, $\overline{p}_k$, and $\overline{\pi}_k$ indicate empirical density approximations of $\gamma_k$, $p_k$, and $\pi_k$, each of which is formed employing a set of $N_P$ particles. Those particles that succeed in 
reaching $Q_k$ from $Q_{k-1}$ form $\Tilde{\pi}_k$. Here, four steps must be taken to estimate the reach probability $\gamma$, including \textit{mutation}, \textit{conditioning}, \textit{selection}, and \textit{splitting}. The \textit{mutation} step consists of executing the SHS $\mathcal{A}^*$, where system equations are evaluated at time $t$ until the next time instant $t_+=\min\{ t+\Delta, \Bar{s}_t, \Bar{\tau}_k\}$, in which $\Delta$ is a small time step, $\Bar{s}_t$ is the first time $>t$ that the solution hits the boundary of $X^*$, and $\Bar{\tau}_k$ is the first time that the solution hits $Q^*_k = Q_k\times \R$. This evaluation is repeated until it hits the next level set $Q_k$ and the successful particles are collected. This execution is outlined in Algorithm \ref{alg:Algorithm 1}. The \textit{conditioning} step is calculating the ratio of successful particles $N_{S_k}$ reaching $Q_k$ to $N_P$ particles, resulting the reach probability $\Bar{\gamma}_k$ which is zero if $N_{S_k}=0$. In the \textit{selection} step, the successful particles are selected to be used in the \textit{splitting} step, which is copying each of the $N_{S_k}$ successful particles as extensively as feasible. The approach used in \textit{splitting} step is the fixed assignment splitting, which consists of two steps. In Step I, each particle is copied $\lfloor N_P/N_{S_k} \rfloor$ times, while in Step II, the remaining $N_P - \lfloor N_P/N_{S_k} \rfloor N_{S_k}$ particles are chosen randomly (without replacement) from $N_{S_k}$ particles and added to the ones from Step I. Then, these steps are repeated until $\Bar{\gamma}_k,\; \forall k \in \{1, \dots, m\}$ are 
obtained, and ultimately, the estimated reach probability $\Bar{\gamma}$ is calculated. These steps are outlined in Algorithm \ref{alg:Algorithm 2}, known as IPS-based estimation with fixed assignment splitting (IPS-FAS).
\IncMargin{1em}
\LinesNumberedHidden

\begin{algorithm}[h!]
	\caption{\label{alg:Algorithm 1}
		The execution function of SHS}
	\DontPrintSemicolon
	\SetKwInOut{Input}{\hspace{0.2cm}Input}\SetKwInOut{Output}{Output}
	\SetKwFunction{Exec}{Execute}
	\SetKwProg{Fn}{Function $\Bar{\xi}_k$ = }{:}{end}
	\SetFuncArgSty{}
	\Input{\justifying{Particle vector $\xi_{k-1}=(\tau_{k-1}, \theta^{*}_{\tau_{k-1}},\mathbf{x}^{*}_{\tau_{k-1}},q^{*}_{\tau_{k-1}})$, SHS elements $(\Theta^*, d^*, X^*, f^*, g^*, Init^*,R^* )$, and $Q^*_k = Q_k \times \R$}}
	\ 
	\Output{Estimated particle $\Bar{\xi}_k=(\Bar{\tau}_k, \Bar{\theta}^{*}_{\Bar{\tau}_k}, \Bar{\mathbf{x}}^{*}_{\Bar{\tau}_k}, \Bar{q}^{*}_{\Bar{\tau}_k})$}
	\algrule
	\Fn{\Exec{$\xi_{k-1}$}}{
		\setcounter{AlgoLine}{-1}
		\nl Set $t\Let \tau_{k-1}$ and $\Bar{\varsigma}\Let (\theta^{*}_{\tau_{k-1}},\mathbf{x}^{*}_{\tau_{k-1}},\Bar{q})$, with $\Bar{q}\sim\exp(1)$\;
		\setcounter{AlgoLine}{0}
		\nl Evaluate equation \eqref{eq:vehicle model} for the AVs and $\mathsf{d}q_t/\mathsf{d}t=-\lambda({\theta}_t, \mathbf{x}_t)$ from $\Bar{\varsigma}$ at $t$ until $t_+=\min\{ t+\Delta, \Bar{s}_t, \Bar{\tau}_k\}$; this yields $\Bar{\varsigma}_+$ \label{step2_2}
		\; 
		\setcounter{AlgoLine}{1}
		\nl \If{$t_+ \ge \Bar{\tau}_k$}{ $\Bar{\xi}_k=(\Bar{\tau}_k,\Bar{\theta}_{\Bar{\tau}_k}^{*},\Bar{\mathbf{x}}_{\Bar{\tau}_k}^{*}, \Bar{q}_{\Bar{\tau}_k}^{*})$, where\\
			\eIf{$\Bar{s}_t=\Bar{\tau}_k$}{$(\Bar{\theta}_{\Bar{\tau}_k}^{*},\Bar{\mathbf{x}}_{\Bar{\tau}_k}^{*}, \Bar{q}_{\Bar{\tau}_k}^{*})\sim R^*(\Bar{\varsigma}_+, \cdot)$}{$(\Bar{\theta}_{\Bar{\tau}_k}^{*},\Bar{\mathbf{x}}_{\Bar{\tau}_k}^{*}, \Bar{q}_{\Bar{\tau}_k}^{*}) \Let  \Bar{\varsigma}_+$}}\; 
		\setcounter{AlgoLine}{2}
		\nl \If{$t_+ \ge \Bar{s}_t$}{$\Bar{\varsigma} \sim R^*(\Bar{\varsigma}_+, \cdot)$, set $t \Let t_+$ and repeat from Step \ref{step2_2}}\;} 
\end{algorithm}
\DecMargin{1em}
For the sake of better illustration of the underlying concept and technicality, we present a running case study which utilizes the model of a vehicle
for AVs $\mathcal{E}$. 

\noindent\textbf{Running Case Study.}
	We consider the following $5$D model, adapted from \cite{pepy2006path}, for each ego vehicle $i\in\mathcal{E}$:
	\begin{align}\notag
			\mathsf{d}x_{t, i} & = (v_{x_{i}} \cos (\vartheta_{t, i}) - v_{y_{t, i}} \sin(\vartheta_{t,i}))\mathsf{d}t + \varepsilon_1 \mathsf{d}\mathbb P_t + \varepsilon_2 \mathsf{d}\mathbb{W}_t, \\\notag
			\mathsf{d}y_{t, i} & = (v_{x_{i}} \sin (\vartheta_{t, i}) + v_{y_{t, i}} \cos(\vartheta_{t,i}))\mathsf{d}t + \varepsilon_1 \mathsf{d}\mathbb P_t + \varepsilon_2 \mathsf{d}\mathbb{W}_t, \\\notag
			\mathsf{d}\vartheta_{t, i} & = \omega_{t, i}\mathsf{d}t, \\\notag
			\mathsf{d}v_{y_{t, i}} \!\!& = (\frac{F_{yf}}{m}\cos(u_{t, i}) + \frac{F_{yr}}{m} - v_{x_{i}} \omega_{t, i}) \mathsf{d}t, \\\label{eq:vehicle model}
			\mathsf{d}\omega_{t, i} & = (\frac{L_f}{I_z}F_{yf} \cos(u_{t, i}) - \frac{L_r}{I_z} F_{yr}) \mathsf{d}t,
	\end{align}
	where $x_{t, i}$ and $y_{t, i}$ are the positions of the vehicle's center of gravity in $x$ and $y$ directions, respectively, $\vartheta_{t, i}$ is the vehicle's orientation, $v_{y_{t, i}}$ is the velocity in the $y$ direction whereas $v_{x_i}$ is the constant velocity in the $x$ direction, $\omega_{t, i}$ is the yaw rate, $\mathbb P_t$ is a Poisson process with rate $\lambda_1$ and reset term $\varepsilon_1$, and $\mathbb W_t$ is a Brownian motion with diffusion term $\varepsilon_2$. The only control input is the front wheel steering angle $u_{t, i}$. Note that since this work is concerned with the verification problem, not controller synthesis, this control input is assumed to be already designed and deployed to the vehicle. The primary objective here is to conduct analysis and compute the rare collision risk probability. Incorporating the stiffness coefficients for front and rear tires $C_{\alpha f}$ and $C_{\alpha r}$, respectively, the forces acting on the front and rear tires $F_{yf}$ and $F_{yr}$, assuming a linear tire model, can be expressed as
	$$
	F_{yf} = -C_{\alpha f} \alpha_f, \quad F_{yr} = -C_{\alpha r} \alpha_r,
	$$
	where the two slip angles $\alpha_f$ and $\alpha_r$ are as
	$$
		\alpha_f = \frac{v_{y_{t, i}} + L_f \omega_{t, i}}{v_{x_i}} - u_{t, i}, \quad \alpha_r = \frac{v_{y_{t, i}} - L_r \omega_{t, i}}{v_{x_i}},
	$$
	with $L_f$ and $L_r$ being the distance from the vehicle's center of gravity to the front and rear wheels.

The components of the GSHS model can be determined according to Definition \ref{def:GSHS} as follows:
	\begin{subequations}
		\begin{align}\notag
				f(\theta_{t, i}, \mathbf{x}_{t, i}) = \operatorname{col}\big(&v_{x_{i}} \cos (\vartheta_{t, i}) - v_{y_{t, i}} \sin(\vartheta_{t,i}), v_{x_{i}} \sin (\vartheta_{t, i}) + v_{y_{t, i}} \cos(\vartheta_{t,i}),\omega_{t, i}, \\\label{eq:f sedan}
				&\frac{F_{yf}}{m}\cos(u_{t, i}) + \frac{F_{yr}}{m} - v_{x_{i}} \omega_{t, i},\frac{L_f}{I_z}F_{yf} \cos(u_{t, i}) - \frac{L_r}{I_z} F_{yr}\big),\\\label{eq:g sedan}
			&\hspace{-2.8cm}g(\theta_{t, i}, \mathbf{x}_{t, i})= {\varepsilon_2}\,\operatorname{col}\big(\pmb{1}_{2\times 1}, \pmb{0}_{3 \times 1}\big),
		\end{align}
	\end{subequations}
	where $\mathbf{x}_{t, i} = \operatorname{col}(x_{t, i}, y_{t, i}, \vartheta_{t, i}, v_{y_{t, i}}, \omega_{t, i})$. In order to model $ER$ with continuous states $\mathbf{x}_{t, ER}$
described by \eqref{eq:vehicle model} as a GSHS model, we should first determine the discrete states $\theta_{t, ER}$. 
To this aim, we define $\underline{\theta}_{t, ER}\in\underline{\Theta}_{ER}$ as the \emph{modes of driving} where
\begin{align*}
    \underline{\Theta}_{ER} = \{0,1,2,-1,Hit\},
\end{align*}
in which each component indicates when the AV $ER$
\begin{itemize}
    \item $0$: is moving straight,~$1$: is changing lanes;
    \item $2$: is aware of the other vehicle changing lanes;
    \item $-1$: is changing its decision (changing lanes in the opposite direction, \ie, returning to its previous lane);
    \item $Hit$: collides with the other vehicle.
\end{itemize}
Then, we define $\upsilon_t\in\Upsilon$ which indicates the \emph{intent of the vehicle}, with set $\Upsilon$ being defined as
\begin{align*}
    \Upsilon = \{ Off, 1^+, 1^- \},
\end{align*}
in which each component indicates:
\begin{itemize}
    \item $Off$: when the indicators of the AV is off and it is likely not to change lanes;
    \item $1^+$: when the right indicator of the AV is flashing and it is changing its lane to the corresponding lane;
    \item $1^-$: when the left indicator of the AV is flashing and it is changing its lane to the corresponding lane.
\end{itemize}
Discrete state of the AV $ER$ is 
$\theta_{t, ER} = (\underline{\theta}_{t, ER}, \upsilon_t)\in\Theta_{ER}$, with $\Theta_{ER}$ as follows:
\begin{equation*}
        \Theta_{ER} = \big\{(0, Off), (1, 1^-), (2, 1^-), (-1, 1^+), (Hit, \star)\big\},
\end{equation*}
\!\!where, the element denoted by $\star$ is non-contributory, meaning that its value has no impact on the outcome.
Analogously, we define discrete states $\theta_{t, EL}$ of the other AV. However, since $EL$ is assumed \emph{not to have SA}, the modes $2$ and $-1$ are not applicable for it. Therefore, $\underline{\theta}_{t, EL}\in\underline{\Theta}_{EL}$ with $
    \underline{\Theta}_{EL} = \{0,1,Hit\}$. Hence, $\theta_{t, EL} = (\underline{\theta}_{t, EL},\upsilon_t)\in\Theta_{EL}$, with $\Theta_{EL}$ defining as
\begin{equation*}
    \Theta_{EL} = \big\{(0, Off), (1, 1^+), (Hit, \star)\big\}.
\end{equation*}
Now that the AVs are modelled, we can define MA-SA relations 
for AV $ER$. To this aim, we define the continuous-valued SA state vector as
\begin{equation}\label{eq:continuous SA}
        \Hat{\mathbf{x}}_{t, ER}^{EL} = \operatorname{col}(\Hat{x}_{t, EL}, \Hat{y}_{t, EL}, \Hat{\vartheta}_{t, EL}, \Hat{v}_{y_{t, EL}}),
\end{equation}
and \emph{augment} it with $\mathbf{x}_{t, ER}$, resulting in continuous-valued state vector with SA $z_{t, ER} = \operatorname{col}\big(\mathbf{x}_{t, ER},$ $ \Hat{\mathbf{x}}_{t, ER}^{EL}, \eta_{t, ER}\big)\in\R^{10}$, in which $\eta_{t, ER}$ indicates the amount of time passed since the AV $ER$ becomes aware of the other's intention to change its lane. Thus, $Z_{ER}^{EL}$ is defined as
\begin{align*}
        Z_{ER}^{EL} &= \big\{(6, 1), (7, 2), (8, 3), (9, 4)\big\}= \big\{\Hat{x}_{t, EL}, \Hat{y}_{t, EL}, \Hat{\vartheta}_{t, EL}, \Hat{v}_{t, EL}\big\}.
\end{align*}
Similarly, the augmented \emph{discrete}-valued state vector with SA is
\begin{equation}\label{eq:discrete SA}
    \check{\theta}_{t, ER} = \operatorname{col}\big(\theta_{t, ER}, \Hat{\theta}_{t, ER}^{EL}\big),
\end{equation}
where $\theta_{t, ER}\in\Theta_{ER}$ and $\Hat{\theta}_{t, ER}^{EL}\in\Theta_{EL}$.
Hence, the GSHS model of the AV $ER$ has \emph{hybrid states} $(\check{\theta}_{t, ER}, z_{t, ER})$. The augmented continuous states $z_{t, ER}$ evolve within the switching moments of $\{\check{\theta}_{t,ER}\}$ as
$$
\check{f}(\check{\theta}_{t, ER}, z_{t, ER}) = \operatorname{col}\big(f(\theta_{t,ER}, \mathbf{x}_{t,ER}), \hat{f}(\hat{\theta}_{t, ER}^{EL}, \hat{\mathbf{x}}_{t,ER}^{EL}), 1\big),
$$
where $f(\theta_{t,ER}, \mathbf{x}_{t,ER})$ is as in \eqref{eq:f sedan}, and $\hat{f}(\hat{\theta}_{t, ER}^{EL}, \hat{\mathbf{x}}_{t,ER}^{EL})$ is as follows:
\begin{align}\notag
            \hat{f}&(\hat{\theta}_{t,ER}^{EL}, \hat{\mathbf{x}}_{t,ER}^{EL}) = \operatorname{col} \big(v_{x_{EL}} \!\cos ({\Hat{\vartheta}_{t, EL}}) - {\Hat{v}_{y_{t, EL}}}\!\sin({\Hat{\vartheta}_{t,EL}}), \\\label{eq:hat f Sedan}
            & v_{x_{EL}} \!\sin ({\Hat{\vartheta}_{t, EL}}) + {\Hat{v}_{y_{t, EL}}} \!\cos({\Hat{\vartheta}_{t,EL}}), \omega_{t, EL},\frac{F_{yf}}{m}\cos(u_{t, EL}) + \frac{F_{yr}}{m} - v_{x_{EL}} \omega_{t, EL}\big).
\end{align}
In addition, for hybrid states $(\check{\theta}_{t, ER}, z_{t,ER})$, we define
$$
\check{g}(\check{\theta}_{t,ER}, z_{t, ER}) = \operatorname{col}\big(g(\theta_{t, ER}, \mathbf{x}_{t,ER}), \hat{g}(\hat{\theta}_{t,ER}^{EL}, \hat{\mathbf{x}}_{t,ER}^{EL}), 0\big),
$$
in which $g(\theta_{t, ER}, \mathbf{x}_{t,ER})$ is as in \eqref{eq:g sedan}, and $\hat{g}(\hat{\theta}_{t,ER}^{EL}, \hat{\mathbf{x}}_{t,ER}^{EL}) = \pmb{0}_{4 \times 1}$.

\begin{remark}
        Another discrete-state SA that can be generally considered is the \emph{identity of vehicles}. This information can be obtained as initial data from the object and treated as a time-invariant state, incorporated into the vehicle's decision-making process.
\end{remark}

\begin{figure}[t!]
    \centering
    \includegraphics[width=0.7\linewidth]{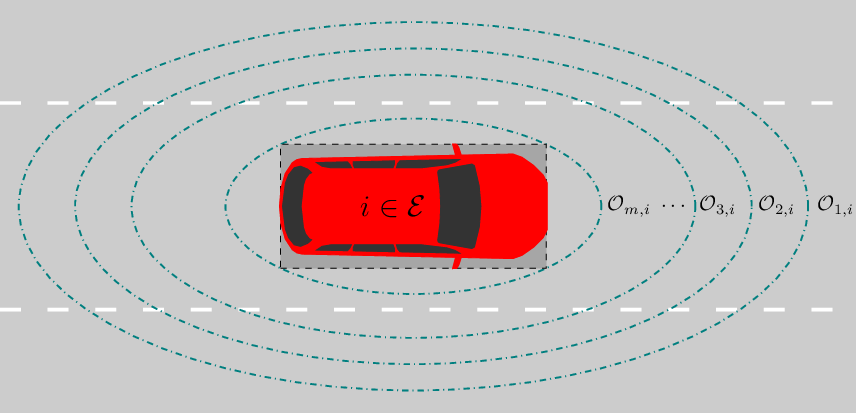}
    \caption{Ellipsoidal level sets $\mathcal O_{k, i}$ as in \eqref{eq:ellipse level sets} around each AV.}
    \label{fig:ellipse level sets}
\end{figure}

Since both vehicles are moving, reaching static level sets $D_k,\,k \in \{0, \dots, m\}$, detailed in Section \ref{sec:IPS}, is not applicable anymore. To deal with this problem, we consider a \emph{set of ellipses} around each AV of the following form
\begin{equation}\label{eq:ellipse level sets}
    \mathcal O_{k, i}\!\Let\! \Big\{\frac{(x \!-\! x_{t,i})^2}{r_{x_k}^2} \!+\! \frac{(y \!-\! y_{t, i})^2}{r_{y_k}^2} \!=\! 1\,\big\vert\, k\!\in\!\{1,\!\dots,\!m\}\Big\},
\end{equation}
where $r_{x_k}$ and $r_{y_k}$ are the primary axes, and $(x_{t, i},y_{t, i})$, with $i\in\mathcal{E}$, is the center of each ellipse. Then, we determine whether $\mathcal O_{k, i} \cap \mathcal O_{k, j} \neq \emptyset$, for $i,j\in \mathcal{E},\, i\neq j$. This demonstrates that the AVs are getting closer to each other and that they might collide. To be more precise, the intersection of the last ellipses, \ie, $\mathcal O_{m, i}\cap \mathcal O_{m, j} \neq \emptyset$, for $i, j\in\mathcal{E},\, i\neq j$, means that the accident has happened. We choose $\mathcal O_{m, i}$ to be a circumscribed ellipse, \ie, the tightest one around the AVs, with primary axes $r_{x_m} = \mathcal R_x = \frac{\sqrt{2}}{2}l_v$ and $r_{y_m} = \mathcal R_y = \frac{\sqrt{2}}{2} w_v$, with $l_v$ and $w_v$ being the length and width of the vehicle, respectively. It is noteworthy that the ellipses are nested subsets within the domain $\Xi$ as well, \ie, $\mathcal O_i \!=\! \mathcal O_{m, i} \subset \mathcal O_{m-1, i} \!\subset\! \cdots \!\subset\! \mathcal O_{1, i} \!\subset\! \mathcal O_{0, i} \!\subseteq\! \Xi$. This setting of level sets is depicted in Fig.~\ref{fig:ellipse level sets}.
	
As long as the AV $EL$ is not close enough to the AV $ER$ so that it can receive the necessary information for situation awareness, we assume the AV $ER$ is not aware of the AV $EL$. To demonstrate this behavior, we consider the ellipse $\mathcal O_{SA, i}$ as defined in \eqref{eq:ellipse level sets} as the area of awareness around each AV with primary axes $\mu_{r_x}$ and $\mu_{r_y}$.
Whenever the awareness ellipses of the two AVs intersect, \ie, $\mathcal O_{SA, i} \cap \mathcal O_{SA, j} \neq \emptyset$, the AV $ER$ becomes aware of the other AV and can receive information for \eqref{eq:continuous SA} and \eqref{eq:discrete SA}.  Upon recognizing the presence of  $EL$, if $\Hat{\theta}_{t, ER}^{EL}=(1, 1^+)$, \ie, $EL$ is changing lane, it will take $ER$ some time to transit to $\theta_{t, ER} = (2, 1^-)$ and decide for its next move. This \emph{delay} can be modeled as an \emph{instantaneous transition rate} $\lambda_2(\check{\theta}_{t, ER}, z_{t, ER})$ which satisfies
	\begin{equation}
		\lambda_2(\check{\theta}, z) = \chi\big(\theta_{t, ER}=(1, 1^-)\big)\, p_{delay}(\eta)/\int_{\eta}^{\infty}p_{delay}(s)\mathsf{d}s,
	\end{equation}
	where $p_{delay}(s)=\frac{s}{\mu_d^2}e^{-s^2/(2\mu_d^2)}$, with the mean reaction delay $\mu_d$ being
	a Rayleigh density.

\section{Time-to-collision Measure}
Upon obtaining data from the SA vector \eqref{eq:continuous SA} of a neighboring vehicle, the ego vehicle must determine its course of action in the event of a potential collision. Time-related measures can be used as a cue for decision making, one of which is time-to-collision (TTC) measure. A shorter TTC indicates a higher risk of collision. 
The TTC for a vehicle
$\alpha$ at a given moment $t$, concerning a preceding vehicle $\alpha - 1$, following the same path, can be computed using
\begin{equation}\label{eq:TTC1}
	\mathrm{TTC}_\alpha = \frac{x_{t, \alpha - 1}-x_{t, \alpha}-l_{\alpha-1}}{v_{t,\alpha} - v_{t,\alpha - 1}},\; \forall v_{t,\alpha} > v_{t, \alpha -1},
\end{equation}
where $l$ is the length of the vehicle \cite{minderhoud2001extended}. Different modifications have been made to \eqref{eq:TTC1} in various studies.
An innovative method is recently proposed in \cite{nadimi2020evaluation} for computing TTC in both car-following and lane-change scenarios by incorporating the equation of motion and vehicle direction. To do so, the category of the collision is firstly identified as either angular or rear-end, with the latter occurring frequently in car-following scenarios. The type of collision is determined by the angle between the movement trajectories of two vehicles, which is represented as a vector whose start and end points are the vehicle's coordinates at the previous and current time instants, respectively. The movement angle of each vehicle can be calculated as provided in Table \ref{tab:angle}.

\begin{table}[t!]
	\setlength{\tabcolsep}{34.5pt}
	\caption{ Calculating the angle of motion for each vehicle.\label{tab:angle}}
	\begin{tabular}{|c|c|}
		\hline
		Condition & Angle $\varphi$\\
		\hline
		$x_2 > x_1$ & \multirow{2}{*}{$\arctan\big|\frac{y_2 - y_1}{x_2 - x_1}\big|$}\\
		$y_2 > y_1$ & {}\\ \cline{1-2}
		$x_2 < x_1$ & \multirow{2}{*}{$\pi - \arctan\big|\frac{y_2 - y_1}{x_2 - x_1}\big|$}\\
		$y_2 > y_1$ & {}\\ \cline{1-2}
		$x_2 < x_1$ & \multirow{2}{*}{$\pi + \arctan\big|\frac{y_2 - y_1}{x_2 - x_1}\big|$}\\
		$y_2 < y_1$ & {}\\ \cline{1-2}
		$x_2 > x_1$ & \multirow{2}{*}{$2\pi - \arctan\big|\frac{y_2 - y_1}{x_2 - x_1}\big|$}\\
		$y_2 < y_1$ & {}\\ \hline
	\end{tabular}
\end{table}

\IncMargin{1em}
\LinesNumberedHidden

\begin{algorithm}[t!]
	\caption{\label{alg:TTC 1}
		Determining the predicted collision point between two vehicles}
	\DontPrintSemicolon
	\SetKwInOut{Input}{\hspace{0.2cm}Input}\SetKwInOut{Output}{Output}
	\Input{\justifying{Current coordinates $(x_{0,\sv}, y_{0,\sv})$ and $(x_{0,\cv}, y_{0,\cv})$ of the subject and colliding vehicles}}
	
	\Output{The common collision point $(c_x,c_y)$, if it exists}
	\algrule
	\setcounter{AlgoLine}{0}
	{\justify \nl Construct the equation of two lines as the motion path of each vehicle using their current position coordinates $(x_{0,\sv}, y_{0,\sv})$ and $(x_{0,\cv}, y_{0,\cv})$:\\
		{\fontsize{8.5}{8.5}\selectfont
			\begin{equation*}
				\begin{aligned}
					y_{t,\sv} &= x_{t,\sv}\tan{\varphi_{\sv}} + (y_{0,\sv} - x_{0,\sv}\tan{\varphi_{\sv}})\\
					y_{t,\cv} &= x_{t,\cv}\tan{\varphi_{\cv}} + (y_{0,\cv} - x_{0,\cv}\tan{\varphi_{\cv}})
				\end{aligned}
			\end{equation*}
	}}
	\setcounter{AlgoLine}{1}
	{\justify \nl Find the intersection of the lines:
		$$
		y_{t,\sv} = y_{t,\cv} \longrightarrow \begin{cases}
			x_{\sv} = x_{\cv} = c_x,\\
			y_{\sv} = y_{\cv} = c_y
		\end{cases}
		$$}
	\setcounter{AlgoLine}{2}
	{\justify \nl Solve the following equations for $t$:
		$$
		{\fontsize{8.5}{8.5}\selectfont
			\begin{aligned}
				&\begin{cases}
					c_x = x_{0,\sv} + \sum\limits_{n=1}^{k}\Big(\dfrac{1}{n!}\times\dfrac{\partial^n x_{t,\sv}}{\partial t^n}\times t^n\Big), & {k \in \N}\\
					c_x = x_{0,\cv} + \sum\limits_{n=1}^{k}\Big(\dfrac{1}{n!}\times\dfrac{\partial^n x_{t,\cv}}{\partial t^n}\times t^n\Big), & {k \in \N}
				\end{cases}\\
				&\begin{cases}
					c_y = y_{0,\sv} + \sum\limits_{n=1}^{k}\Big(\dfrac{1}{n!}\times\dfrac{\partial^n y_{t,\sv}}{\partial t^n}\times t^n\Big), & {k \in \N}\\
					c_y = y_{0,\sv} + \sum\limits_{n=1}^{k}\Big(\dfrac{1}{n!}\times\dfrac{\partial^n y_{t,\cv}}{\partial t^n}\times t^n\Big), & {k \in \N}
				\end{cases}
			\end{aligned}
		}
		$$
	}
	\setcounter{AlgoLine}{3}
	\nl \eIf{{\small {$t\in\R^+$} for each of the four equation exists}}{$(c_x,c_y)$ is the predicted collision point and $\mathrm{TTC}_{\sv}$ can be calculated for this point}{There is no predicted collision point and $\mathrm{TTC}_{\sv} = \infty$}
\end{algorithm}
\DecMargin{1em}

If the angle of a prospective collision, which is the absolute difference between the angles of motion of two vehicles, falls between $-10$ and $+10$ degrees and both vehicles are traveling in the same lane, the conflict is considered a rear-end collision. The necessary and sufficient condition for rear-end collision is as follows:
\begin{equation}\label{eq:rear-end conflict condition}
	x_{t, \alpha} - x_{t, \alpha -1} + l_{\alpha - 1} = 0 \iff \text{Rear-end collision}.
\end{equation}
Assuming the $(k-1)^\mathrm{th}$ derivative of velocity is constant, we can derive an approximate equation of motion for each vehicle as follows:
\begin{equation}\label{eq:derivative cte velocity}
	\begin{cases}
		x_{t,\alpha} = x_{0, \alpha} + \sum\limits_{n=1}^{k}\Big(\dfrac{1}{n!}\times\dfrac{\partial^n x_{t,\alpha}}{\partial t^n}\times t^n\Big),\\
		x_{t, \alpha -1 } = x_{0,\alpha -1} + \sum\limits_{n=1}^{k}\Big(\dfrac{1}{n!}\times\dfrac{\partial^n x_{t,\alpha -1}}{\partial t^n}\times t^n\Big).
	\end{cases}
\end{equation}
Combining \eqref{eq:rear-end conflict condition} and \eqref{eq:derivative cte velocity} results in the $k^\mathrm{th}$ degree polynomial
\begin{equation}\label{eq:TTC solution}
	\begin{aligned}
		x_{0,\alpha} -& x_{0, \alpha-1} + l_{\alpha-1} +\\
		&\sum\limits_{n=1}^{k}\Big(\dfrac{1}{n!}\times\Big[\dfrac{\partial^n x_{t, \alpha}}{\partial t^n}-\dfrac{\partial^n x_{t, \alpha -1}}{\partial t^n}\Big]\times t^n\Big) = 0,
	\end{aligned}
\end{equation}
whose solution is $\mathcal{T} = \{ t_1, t_2,\dots, t_k \}$. Then,
\begin{equation}
	\mathrm{TTC}_k=\min\{ t_i\in\mathcal{T}|t_i\in\R^+\},
\end{equation}
implying that $\mathrm{TTC}_k$ is the minimum, non-zero and real solution of \eqref{eq:TTC solution}.

When dealing with angular collisions, the initial step involves ascertaining whether a subject vehicle ``$\sv$'' and a colliding vehicle ``$\cv$'' share a common collision point, 
at which $\mathrm{TTC}$ can be calculated. In order to compute the common collision point, the motion path of each vehicle is determined by constructing line equations for each based on their calculated angles, as in Step 1 of Algorithm~\ref{alg:TTC 1}. The intersection of these lines results in the point $(c_x, c_y)$, which might
be the common collision point according to~\eqref{common collision}.

Then, the motion type of each vehicle is determined by examining their previous positions in $x$ and $y$ directions at each time instant:
\begin{subequations}\label{common collision}
	\begin{align}
		&\begin{cases}
			x_{t,\sv} \!=\! x_{0,\sv} \!+\!\!\! \sum\limits_{n=1}^{k}\Big(\dfrac{1}{n!}\!\times\!\dfrac{\partial^n x_{t,\sv}}{\partial t^n}\!\times\! t^n\Big), & \!\!\!{k\!\in\!\N,}\\
			y_{t,\sv} \!=\! y_{0,\sv} \!+\!\!\! \sum\limits_{n=1}^{k}\Big(\dfrac{1}{n!}\!\times\!\dfrac{\partial^n y_{t,\sv}}{\partial t^n}\!\times\! t^n\Big), & \!\!\!{k\!\in\!\N,}
		\end{cases}\label{eq:motionEqSub}\\
		&\begin{cases}
			x_{t,\cv} \!=\! x_{0,\cv} \!+\!\!\! \sum\limits_{n=1}^{k}\Big(\dfrac{1}{n!}\!\times\!\dfrac{\partial^n x_{t,\cv}}{\partial t^n}\!\times\! t^n\Big), &\!\!\! {k\!\in\!\N,}\\
			y_{t,\cv} \!=\! y_{0,\cv} \!+\!\!\! \sum\limits_{n=1}^{k}\Big(\dfrac{1}{n!}\!\times\!\dfrac{\partial^n y_{t,\cv
			}}{\partial t^n}\!\times\! t^n\Big), &\!\!\! {k\!\in\!\N.}
		\end{cases}\label{eq:motionEqCol}
	\end{align}    
\end{subequations}

\begin{figure*}[t!]
	\centering
	\begin{subfigure}[b]{0.47\textwidth}
		\centering
		\includegraphics[width=0.87\linewidth]{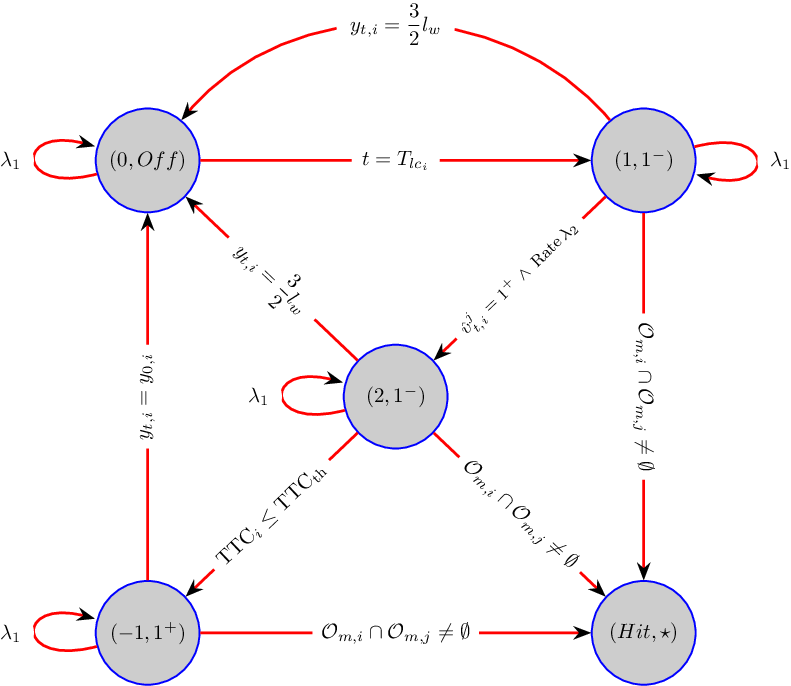}
		\caption{}
		\label{fig:GSHS ER}
	\end{subfigure}
	\hfill
	\begin{subfigure}[b]{0.47\textwidth}
		\centering
		\includegraphics[width=0.87\linewidth]{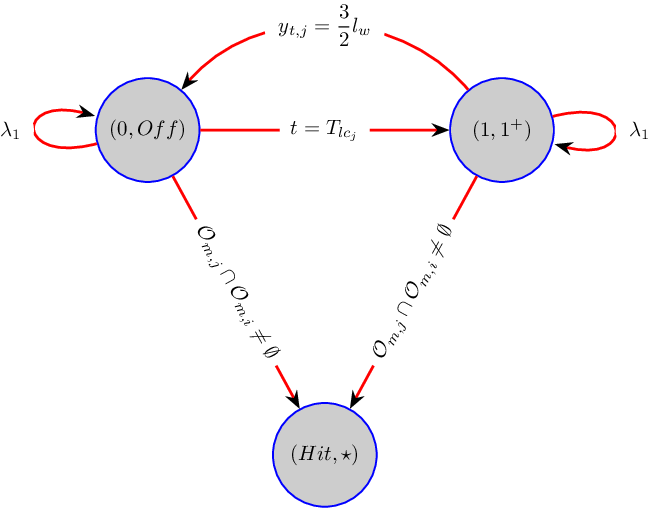}
		\caption{}
		\label{fig:SHS EL}
	\end{subfigure}
	\caption{GSHS model transition graphs for AVs $i=ER$ (a), and $j = EL$ (b), where, $w_L$ represents lane width, and $T_{{lc}_i}$, $T_{{lc}_j}$ signify moments when vehicles $i$ and $j$ decide to change lanes, respectively. The intent $\hat{\upsilon}_{t,i}^j$ is obtained from $\hat{\theta}_{t,i}^j$.}
	\label{fig:Hybrid}
\end{figure*}

We solve \eqref{eq:motionEqSub} and \eqref{eq:motionEqCol} at the point $(c_x, c_y)$ for $t$ and then examine the solutions. If $t\in\R^+$, then the point $(c_x,c_y)$ is considered as the common collision point. The necessary steps to specify the collision point are given in Algorithm \ref{alg:TTC 1}. Then, if a common collision point exists, $\mathrm{TTC}$ for the subject vehicle can be determined. To this aim, the distance $d_{\sv}$ between the subject vehicle and the collision point $(c_x, c_y)$ resulting from Algorithm \ref{alg:TTC 1} is calculated. Then, we calculate the time it takes the subject vehicle to drive this distance, either in the $x$ or $y$ direction. This results in a set of solutions, $\mathcal{T}=\{t_1, t_2, \dots,t_k\}$, with $k$ being the order of the motion equation, for which we check whether all $t$ in $\mathcal{T}$ are \emph{real and positive}. If this condition is satisfied, the minimum $t$ in $\mathcal{T}$ is $\mathrm{TTC}$; otherwise,  no collision will occur. This procedure is outlined in Algorithm \ref{alg:TTC 2}. 
In the process of $\mathrm{TTC}$ computation, the subject vehicle utilizes the information provided by the SA vector \eqref{eq:continuous SA}.

\IncMargin{1em}
\LinesNumberedHidden

\begin{algorithm}[t!]
	\caption{\label{alg:TTC 2}
		$\mathrm{TTC}$ calculation for angular conflicts}
	\DontPrintSemicolon
	\SetKwInOut{Input}{\hspace{0.2cm}Input}\SetKwInOut{Output}{Output}
	\Input{\justifying{Predicted collision point for the subject vehicle, \ie, $(x_{t,\sv}, y_{t,\sv}) = (c_x,c_y)$ and its current location $(x_{0, \sv}, y_{0,\sv})$}} 
	
	\Output{$\mathrm{TTC}_{\sv}$ for the subject vehicle}
	\algrule
	\setcounter{AlgoLine}{0}
	{\justify \nl Calculate the distance between the current location of the subject vehicle and the predicted collision point:
		$$
		d_{\sv} = \sqrt{(c_x - x_{0,\sv})^2 + (c_y - y_{0,\sv})^2}
		$$}\
	\setcounter{AlgoLine}{1}
	{\justify \nl Calculate time to the predicted collision point based on the type of motion:
		\begin{equation*}
			\begin{array}{lc}
				& d_{\sv} = \dfrac{\sum\limits_{n=1}^{k}\Big(\dfrac{1}{n!}\times\dfrac{\partial^n x_{t,\sv}}{\partial t^n}\times t^n\Big)}{\cos{\varphi_{\sv}} ~\text{or} ~ \sin{\varphi_{\sv}}}
			\end{array}
	\end{equation*}}
	\setcounter{AlgoLine}{2}
	\nl \eIf{$\exists i \in \{1,\dots, k\},\: t_i \notin \R^+$}{$\mathrm{TTC}_{\sv} = \infty$}{$\mathrm{TTC}_{\sv} = \min\{ t_i \in \mathcal{T} | t_i \in \R^+ \}$}
\end{algorithm}
\DecMargin{1em}

\noindent\textbf{Running Case Study (cont.)}
When the AV $ER$ is in making decision mode $(2,1^-)$, indicating awareness of another AV changing lanes, it calculates the TTC measure to determine whether to complete its maneuver or change its decision and return to its own lane (transition to mode $(-1,1^+)$). We consider a threshold $\mathrm{TTC}_\mathrm{th}$
so that if $\mathrm{TTC}_{ER} \leq \mathrm{TTC}_\mathrm{th}$, completing the maneuver is hazardous and the AV $ER$ will go back to its own lane. 
The transition graphs of the completed GSHS models for AVs $i=ER$ and $j = EL$ are provided in Figs. \ref{fig:GSHS ER} and \ref{fig:SHS EL}, respectively.

\section{Results and Discussions}
In order to utilize Algorithm \ref{alg:Algorithm 2}, we first need to define the level sets as described in \eqref{eq:ellipse level sets}. We assume each AV has six ellipses around it with the primary axes $r_{x_k} = r_k \mathcal R_x$ and $r_{y_k} = r_k \mathcal R_y$ for all $k \in \{1, \dots, 6\}$, where $r_1 =2$ and the declining rate is $0.2$, leading to $r_6 = 1$. The parameters of the AVs described by \eqref{eq:vehicle model} are set as $v_{x_i} = 20 \, m/s$, $\varepsilon_1 = 10^{-6}$, $\varepsilon_2 = 10^{-2}$, $\lambda_1 = 0.5$, $m = 2000\, kg$, $I_z = 2000\, kgm^2$, $C_{\alpha f} = C_{\alpha r} = 6\times 10^4$, $L_f = L_r = 2\, m$, $l_v = 4.508\, m$, and $w_v = 1.61\, m$. To perform a lane-change maneuver, we utilize a simple PD controller of the form $u_{t,i} = K_p\big(y_{d, i} - y_{t, i}\big) -K_d \frac{\mathsf{d}y_{t,i}}{\mathsf{d}t}$ with $K_p = 1.5\times 10^{-3}$, $K_d = 10^{-2}$, and $y_{d, i}$ is the desired position in the $y$ direction. In the scenario under study, we set $w_L = 3.5\, m$, $\mu_d = 0.6\,s$, and $\mathrm{TTC}_{\mathrm{th}} = 10\, s$.

Our aim is to analyze the effect of the area of awareness $\mathcal O_{SA, ER}$, based on the different values of $\mu_r$ in $\mu_{r_x} = \mu_r \mathcal R_x$ and $\mu_{r_y}  = \mu_r \mathcal R_y$. To increase the reliability of the outcomes, we run the scenario $\mathcal N$ times and get the results $\Bar{\gamma}_n, \, n\in\{1, \dots, \mathcal N\}$.
\begin{table*}[t!]
	\centering
	\caption{The value of mean probability $\hat \gamma$ for various $\mu_r$ using Algorithm \ref{alg:Algorithm 2} and MC (m = 1).\label{tab:results}}
		\begin{tabular}{lccccc}  
			\toprule
			\multirow{2}*{Algorithm} & 
			\multicolumn{5}{c}{$\mu_r$} \\
			\cmidrule(l){2-6}
			{}             & $1.5825$ & $1.6275$ & $1.6725$ & $1.7$ & $1.7375$ \\
			\midrule
			IPS-FAS &  $1.9131 \times 10^{-4}$    & $8.6350\times 10^{-5}$ &  $7.6300 \times 10^{-6}$  & $2.8725 \times 10^{-6}$ & $5.4320 \times 10^{-7}$   \\
			MC      &   $1.8000\times 10^{-4}$   &   $7.9000\times 10^{-5}$    &     $0$     &    $0$    &     $0$     \\
			\bottomrule
		\end{tabular}
\end{table*}
Then, we report the \emph{mean probability} $\hat \gamma = \frac{\sum_{n = 1}^{\mathcal N} \Bar{\gamma}_n}{\mathcal N}$ as the estimated probability of reaching $\mathcal O_{6, ER} \cap \mathcal O_{6, EL}\neq \emptyset$. 
We report our obtained results in Table \ref{tab:results} with $\mathcal N = 100$ trials and $N_P = 100$ particles for Algorithm~\ref{alg:Algorithm 2} and the corresponding Monte-Carlo (MC) simulation for the sake of comparison.

The highlights of simulation results can be considered threefold, which are given below:
\begin{itemize}
	\item The scenario's parameters are considered in a way that an accident occurs in the absence of SA;
	\item Given that an accident occurs in this scenario due to the lack of SA, it becomes evident how SA plays a crucial role in reducing accident risk. Furthermore, Table \ref{tab:results} illustrates that even minor adjustments in SA parameters can significantly affect collision probabilities;
	\item Finally, Table \ref{tab:results} underscores the superiority of the IPS-FAS algorithm over MC simulation. While MC yields a zero probability outcome, IPS-FAS provides a probability on the order of $10^{-7}$, highlighting its precision. Given that AVs belong to safety-critical systems, the precision of calculations within their decision-making is of vital importance.
\end{itemize}

\bibliographystyle{IEEEtran}
\bibliography{biblio}

{\small \IncMargin{1em}
	\LinesNumberedHidden
	\begin{algorithm}[t!]
		\caption{\label{alg:Algorithm 2}
			IPS-FAS algorithm for a GSHS}
		\DontPrintSemicolon
		\SetKwInOut{Input}{\hspace{0.2cm}Input}\SetKwInOut{Output}{Output}
		\Input{\justifying{Initial measure $\pi_0$, end time $T$, decreasing sequence of closed subsets $D_k = \{(\theta_t, \mathbf{x}_t) \in \Xi\}$, $
				D_{k-1} \supset D_k,\: k \in \{1,\dots,m\}$. Also $D_0 = \Xi,\; Q_k = (0, T)\times D_k$ and number of particles $N_P$}}
		\Output{Estimated reach probability $\Bar{\gamma}$}
		\algrule
		\setcounter{AlgoLine}{-1}
		\nl\textbf{Initiation:} Generate $N_P$ particles $\xi_0^i \sim \pi_0, \; i \in \{1, \dots, N_P\}$, \ie $\: \Bar{\pi}_0(\cdot) = \sum_{i = 1}^{N_P} \frac{1}{N_P}\delta_{\{\xi_0^i\}}(\cdot)$, with Dirac $\delta$. Set $k = 1$\;
		\setcounter{AlgoLine}{0}
		\SetKwFor{For}{for}{do}{end}
		\SetKwFunction{Exec}{Execute}
		\nl\textbf{Mutation (Algorithm~\ref{alg:Algorithm 1}):}
		\For{$i = 1,\dots, N_P$}{$\Bar{\xi}_k^i=$ \Exec{$\xi_{k-1}^i$}}
		Then, $\Bar{p}_k(\cdot) = \sum_{i=1}^{N_P}\frac{1}{N_P}\delta_{\{\Bar{\xi}_k^i\}}(\cdot)$\label{step2}\;
		\setcounter{AlgoLine}{1}
		\nl\textbf{Conditioning:} $\Bar{\gamma}_k = \frac{N_{S_k}}{N_P}$ with $N_{S_k} = \sum_{i = 1}^{N_P}1(\Bar{\xi}_k^i \in Q_k)$\\
		\If{$N_{S_k} = 0$}{$\Bar{\gamma}_{k^\prime} = 0, \; k^\prime\in\{k,\dots,m\}$ and go to Step \ref{step6}} 
		\setcounter{AlgoLine}{2}
		\nl\textbf{Selection:} $\Tilde{\pi}_k(\cdot) = \frac{1}{N_{S_k}}\sum_{i=1}^{N_{S_k}}\delta_{\{\Tilde{\xi}_k^i\}}(\cdot)$, with $\{\Tilde{\xi}_k^j\}_{j = 1}^{N_{S_k}}$ the collection of $\Bar{\xi}_k^i\in Q_k,\; i \in \{1,\dots,N_P\}$\;
		\setcounter{AlgoLine}{3}
		\nl\textbf{Splitting:} $\{\Tilde{\Tilde{\xi}}_k^j\}_{j = 1}^{N_{S_k}}$ is a random permutation of $\{\Tilde{\xi}_k^j\}_{j = 1}^{N_{S_k}}$\\
		\SetKwFor{For}{for}{copy}{end}
		\For{$i = 1,\dots,N_{S_k}$}{
			$$
			\text{Step I} : \left\{\begin{array}{lcc}
				\xi_k^{i} & = & \Tilde{\Tilde{\xi}}_k^i\\
				\xi_k^{N_{S_k}+i} & = & \Tilde{\Tilde{\xi}}_k^i\\
				\vdots                &   \vdots    &    \\
				\xi_k^{(\lfloor N_P/N_{S_k}\rfloor-1)N_{S_k}+i} & = & \Tilde{\Tilde{\xi}}_k^i\\               
			\end{array}\right.
			$$
		}
		\For{$i = 1,\dots,N_P - \lfloor N_P/N_{S_k}\rfloor N_{S_k}$}{
			$$
			\text{Step II}  :  \begin{array}{lcc}
				\xi_k^{\lfloor N_P/N_{S_k}\rfloor N_{S_k}+i} & \hspace{0.72cm}= & \Tilde{\Tilde{\xi}}_k^i
			\end{array}
			$$
		}
		Each particle receives weight $1/N_P$\;
		\setcounter{AlgoLine}{4}
		\nl\label{step6} \eIf{$\Bar{\gamma}_k\neq 0$}{
			\eIf{$k<m$,}{$k \Let  k + 1$ and go to Step \ref{step2}} 
			{$\Bar{\gamma} = \prod_{k=1}^m \Bar{\gamma}_k$}
		}
		{$\Bar{\gamma}= 0$}
	\end{algorithm}
	\DecMargin{1em}}
\end{document}